\documentclass[aip,twocolumn,reprint,floatfix]{revtex4-1}
\usepackage{amsmath,amssymb,graphicx,psfrag}

\renewcommand{\L}{{\mathfrak L}}

\newcommand{\B}{{\mathfrak B}}
\newcommand{\C}{{\mathfrak C}}
\newcommand{\D}{{\mathfrak D}}
\newcommand{\0}{{\mathfrak 0}}
\newcommand{\g}{{\mathfrak g}}

\begin{document}

\title{Implicit operators for networked mechanical and thermal systems
   with integer-order components}

\author{Mihir Sen}
\affiliation{Department of Aerospace and Mechanical Engineering,
   University of Notre Dame, Notre Dame, IN 46556} \author{John P. Hollkamp} \affiliation{Department of Mechanical Engineering, Purdue University,
   West Lafayette, IN 47907}
\author{Fabio Semperlotti}
\affiliation{Department of Mechanical Engineering, Purdue University,
   West Lafayette, IN 47907}
\author{Bill Goodwine}
\affiliation{Department of Aerospace and Mechanical Engineering,
   University of Notre Dame, Notre Dame, IN 46556}

\date{\today}

\begin{abstract}
Complex systems are composed of a large number of simple components connected to each other in the form of a network. It is shown that, for some network configurations, the equivalent dynamic behavior of the system is governed by an \textit{implicit} integro-differential operator even though the individual components themselves satisfy equations that use explicit operators of integer order. The networks considered here are infinite trees and ladders, and each is composed only of two types of integer-order components with potential-driven flows that are repeated \textit{ad infinitum}. In special cases the equivalent operator for the system is a fractional-order derivative, but in general it is implicit and can only be expressed as a solution of an operator equation. These implicit operators, which are a generalization of fractional-order derivatives, play an important role in the analysis and modeling of complex systems.
\end{abstract}

\maketitle

\section{Introduction}

Implicit functions and the conditions under which they exist are familiar \cite{Krantz2002}. Fractional-order derivatives have also been around for some time, and have found applications in topics like viscoelasticity, mass diffusion, and fractal media, among others \cite{Tarasov2010, West2016}. In this work \textit{implicit
   integro-differential operators} are defined in a manner similar to implicit functions and used for the direct mathematical modeling of complex systems. In this context, fractional derivatives become a subset of implicit operators, and mathematical modeling of complex systems can go beyond the use of fractional calculus. It is shown that these implicit operators appear naturally in the modeling of networked mechanical and thermal systems. The components of these systems are bound by the laws of mechanics and thermodynamics which are usually integro-differential of integer order. However, it has sometimes been mistakenly assumed that the larger system is also of integer order. Thus implicit operators, and their subset the fractional-order derivatives, are a basic tool for the modeling of complex systems.

There is an analogy between real numbers and operators that will be used. An equation such as $f(x)=0$, where $f: \mathbb R \to \mathbb R$ is a known function, may have solutions for which $x$ is real; furthermore, the integer, rational, irrational or transcendental nature of $x$ is also determined by the form of $f$. A number $x$ multiplied by itself $n$ times, where $n$ is an integer, is written as $x^n$, but on choosing $f(x)$ suitably this can be extended to $x^\alpha$ for any real $\alpha$; the law of indices, $x^\alpha x^\beta = x^{\alpha + \beta}$ for any $\alpha$ and $\beta$, holds, but $\alpha$ in $x^\alpha$ no longer represents the number of times $x$ is multiplied by itself. Finally, under appropriate conditions an equation such as $g(y,x) = 0$ with $g : {\mathbb R} \times {\mathbb R} \to {\mathbb R}$ defines an implicit function $y = f(x)$.

Similarly, consider operators $\D$ and $\L$ which operate on a real function $u(t)$, where $t$ is a time-like independent variable (scalar variables will be written in Roman italic, real constants in Greek, matrices in Calligraphic, and integro-differential operators in Fraktur), and where $\D = d/dt$ is the usual first derivative operator. Both operators are assumed to return quantities to which they can be re-applied. Notation-wise ${\L}^m$ and $\D^n$, where $m$ and $n$ are integers, denote the result of $m$ and $n$ repeated applications of $\L$ and $\D$, respectively. As special cases, $\D^n$ is the identity operator if $n=0$, is a derivative operator if $n>0$, and is an integral operator if $n<0$. In some respects the approach here will parallel that of operational calculus \cite{Erdelyi1962} in which $\D$ is manipulated as an algebraic variable.

The solution of an operator equation
\begin{equation}
\label{def}
\g(\L,\D)=\0,
\end{equation}
where $\0$ is the zero operator, should give a similar variety of operators that $g(y,x) = 0$ did for numbers. If $\L - a \D^n = \0$, where $n$ is an integer, then $\L = a \D^n$. Or take $\L^m - a \D^n = 0$, where $m$ is another integer, for which symbolically $\L = p_i \D^{n/m}~(i=1,2,\ldots,m)$, where $p_1, p_2, \ldots, p_m$ are the $m^\textit{th}$ roots of $a$. This defines $\D^\alpha$ where $\alpha$ is rational. As examples, a semi-derivative operator $\L$ can be defined by $\L^2 - \D = \0$ and a semi-integral by $\L^2 - \D^{-1} = \0$; symbolically $\L = \pm \D^{1/2}$ and $\L = \pm \D^{-1/2}$, respectively. A more general $\g(\L,\D) = \0$ enables $\D^\alpha$ to be defined where $\alpha$ is a real number. The linear operator $\D^\alpha~(\alpha < 0)$ can be written in explicit closed form.  An example is \begin{equation} \label{eq:caputo} \D^\alpha u(t) = \frac{1}{\Gamma\left(n-1\right)} \int_a^t \left(t-\tau\right)^{n-\alpha-1}  \frac{d^n u}{d \tau^n}(\tau)d\tau \end{equation} which is the Caputo definition of the fractional derivative. There are other definitions as well such as Riemann-Liouville, Gr\"{u}nwald-Letnikov, and others \cite{Tarasov2010,
   West2016}. $\D^\alpha$ ($\alpha
 > 0$) can be obtained by differentiating and using the law of indices $\D^\alpha \D^\beta = \D^{\alpha + \beta}$. Since $\D^\alpha$ can be explicitly defined, linear combinations $\sum_i a_i \D^{\alpha_i}$ are also explicit. Although $\D^\alpha$ can also be defined for complex $\alpha$, for simplicity that will be excluded here.

Eq.~\ref{def} enables operators $\L$ to be defined that are not necessarily linear combinations of $\D^\alpha$, except in special cases. In general Eq.~\ref{def} cannot be ``solved'' for $\L$ nor can it be written in the form $\L = \sum_i a_i \D^{\alpha_i}$; rather, $\L$ is \textit{defined} by this equation, and is thus an implicit operator. This is in line with the adjective being used in mathematics with functions, differentiation, integration, differential equations, and so on (implicit methods in numerical algorithms are a slightly different but related use of the term). The equivalent order of an implicit integro-differential operator can be taken to be the ratio between the highest powers of $\D$ and $\L$, respectively, in $\g(\L,\D)$. Although $\L$ cannot be explicitly written, it is nonetheless appropriately though not uniquely defined by Eq.~\ref{def}.

One example of an implicit operator $\L$ that will be used later is when it is defined by a quadratic equation \begin{equation}
\L^2 + \B (\D) \L + \C (\D)= \0. \label{eq:quad1} \end{equation} The solution \[ \L = \frac{1}{2} \left[- \B \pm \left( \B^2 - 4 \C \right)^{1/2}  \right] \] is symbolic since it cannot be used to write $\L$ explicitly; $\L$ is simply such that Eq.~\ref{eq:quad1} is satisfied. Of course there are special cases for which an explicit $\L$ can indeed be found. For example, if $\B^2 = 4 \C$ then $\L = - \B/2$, or if $\B^2 - 4 \C = \D^m$ then $\L = (- \B \pm \D^{m/2})/2$.

Sometimes a polynomial $\g(\L,\D)$ can be factorized, analytically or numerically, so that Eq.~\ref{def} can be written as \[
(\L-\C_1 \D^{\alpha_1})(\L-\C_2 \D^{\alpha_2})  \ldots  (\L-\C_n
\D^{\alpha_n}) = \0
\]
for which the solutions are $\L = \C_1 \D^{\alpha_1}$, $\C_2 \D^{\alpha_2}$, $\ldots$, $\C_n \D^{\alpha_n}$. Each one of these operators is then explicitly definable. The multiplicity of operators $\L$ satisfying Eq.~\ref{def} is inherent in the equation, and whether all or only some are physically valid depends on the derivation of the equation from the underlying physics.

\section{Infinitely self-repeating configurations}

It is important to give examples where implicit operators appear in physical systems, because otherwise the subject would be somewhat sterile. It will be shown that some complex systems with self-repeating components can be modeled by implicit operators. In the following, network configurations or graphs will be considered that are made up of components (also called branches, edges, arcs or lines) that are joined at junctions (nodes, vertices, or points). Through each component there is a flow $u(t)$ that is driven by a difference in potential between its terminals $\Delta \phi(t)$. At each junction it will be assumed that the sum of all incoming flows is zero. These are generalized versions of Kirchhoff's circuit laws that are common in network transport analysis.

In each component $\L$ is a linear operator that can be identified depending on the physics of the problem as illustrated in the following examples. The governing equation in each component then takes the form \begin{equation}
\L(u) = \Delta \phi.
\label{potdr}
\end{equation}

\begin{enumerate}
\item For passive electrical components like a resistor $R$, a
   capacitor $C$ or an inductor $L$ in series \begin{subequations} \label{oper} \begin{equation} \left[ L \D + R \D^0 +  \frac{1}{C}~\D^{-1} \right] (u) =  \Delta \phi, \end{equation} where $u(t)$ is the current and $\Delta \phi (t)$ the driving voltage difference.

\item Water flow in a pipe can be modeled by \begin{equation} \left[ \D + a \D^0 \right] (u) = b~\Delta \phi, \end{equation} where $u(t)$ is the volume flow rate, $a \D^0 u(t)$ is the laminar viscous force, and $\Delta \phi (t)$ is the driving pressure difference.

\item If there is heat conduction along a rod, then \begin{equation} \left[ \D + a \D^0 \right] (u)= b \Delta \phi \end{equation} where $u(t)$ is the spatial average of the temperature, $a \D^0 u$ is the convective loss of heat to the environment, and $\Delta \phi (t)$ is the driving heat rate difference. Usually, the driving potential in heat transfer is the temperature and the flow is the heat rate, but here the roles are reversed.

\item For a spring-damper system
\begin{equation}
\left[ c \D + k \D^0 \right] (u) = \Delta \phi , \end{equation} \end{subequations} where $u(t)$ is the displacement, $c$ is the damping, $k$ is the spring constant, and $\Delta \phi (t)$ is the external driving force.
\end{enumerate}

Component Eqs.~\ref{oper} are all of the form of Eq.~\ref{potdr}. The tree and ladder configurations below are basically one-dimensional flows, and multi-dimensional configurations such as general networks will not be considered here. In the accompanying figures, the straight lines are components and the dots are junctions. The flows are from left to right and they are driven by an overall potential difference $\phi_{in}-\phi_{out}$. To simplify the analysis there are only two component operators, $\L_a$ and $\L_b$, that are repeated infinitely. The flow can be written as $\L_{eq}(u) = \phi_{in}-\phi_{out}$, where $u(t)$ is the flow that results from an overall driving potential $\phi_{in}-\phi_{out}$, and the objective is to find the equivalent operator of the system $\L_{eq}$.

\subsection{Infinite bifurcating trees}

\begin{figure}
\centering
\psfrag{p1}{$\phi_{in}$}
\psfrag{p2}{$\phi_{out}$}
\psfrag{La}{$\L_a$}
\psfrag{Lb}{$\L_b$}
\includegraphics[width=3.25in]{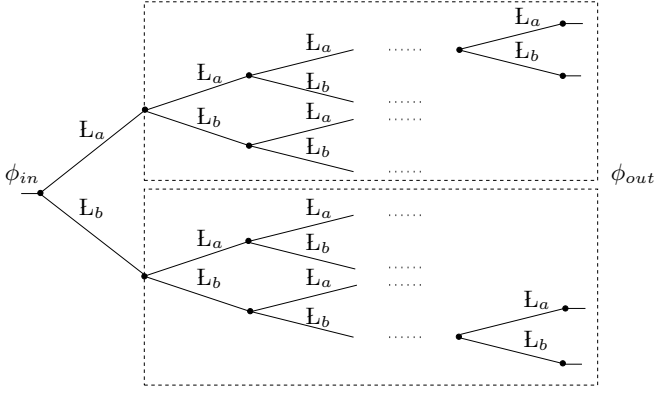}
\caption{Infinite tree composed of operators $\L_a$ and $\L_b$;
   components in boxed area identical to whole.} \label{fig:tree1} \end{figure}

This was investigated first for electrical components \cite{Nakagawa92} and then for flows \cite{Franco2006, Mayes2011} and controlled robots \cite{Goodwine2014}. In Fig.~\ref{fig:tree1}, $\L_a$ and $\L_b$ are known operators for the left and right components, respectively. The three equations below \begin{subequations} \begin{align} (\L_a + \L_{eq})(u_a) &= \phi_{in} - \phi_{out} ,\label{brancha} \\ (\L_b + \L_{eq})(u_b) &= \phi_{in} - \phi_{out} , \label{branchb} \\
\L_{eq}(u_a+u_b) &= \phi_{in} - \phi_{out} , \label{branchc} \end{align} \end{subequations} correspond to the left path, the right path, and the equivalent network, respectively, and $u_a$ and $u_b$ are the flows in the respective components.

\begin{figure}
\psfrag{p1}{$\phi_{in}$}
\psfrag{p2}{$\phi_{out}$}
\psfrag{La}{$\L_a$}
\psfrag{Lb}{$\L_b$}
\psfrag{L}{$\L_{eq}$}
\includegraphics[width=1.5in]{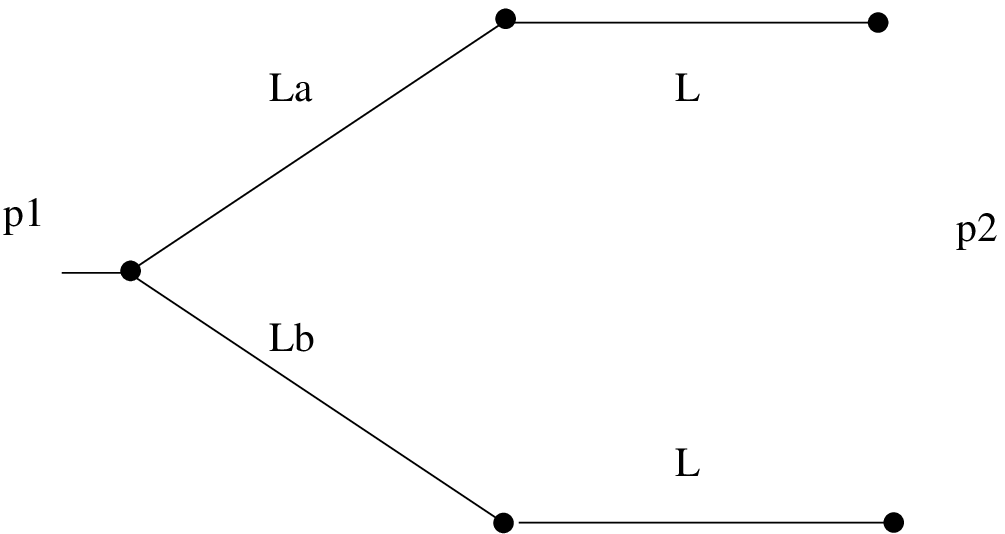}
\caption{Components in boxed area in Fig.~\ref{fig:tree1} replaced by
   equivalent.}
\label{fig:tree2}
\end{figure}

Operating on Eq.~\ref{brancha} by $\L_{eq} (\L_b + \L_{eq})$, on Eq.~\ref{branchb} by $\L_{eq} (\L_a + \L_{eq})$, and adding \begin{multline*} (\L_a + \L_{eq})\ (\L_b + \L_{eq})\ \underbrace{\L_{eq} (u_a +
     u_b)}_{= \phi_{in} - \phi_{out}} \\ = \L_{eq} \left[ \L_a + \L_b +
     2\L_{eq}) \right] (\phi_{in} - \phi_{out}) .
\end{multline*}
Because this must be true for all  $(\phi_{in} - \phi_{out})$, $\L_{eq}$ must satisfy \begin{equation}
\L_{eq}^2 - \L_a \L_b = \0 , \label{semi} \end{equation} where Eq.~\ref{branchc} has been used and some terms canceled. This is of the form of Eq.~\ref{eq:quad1} where $\B(\D)=\0$ and $\C(\D)= -\L_a \L_b$ and can be symbolically solved as $\L_{eq} = \sqrt{ \L_a \L_b }$. Only the positive root is considered since the negative was spuriously introduced when Eqs.~\ref{brancha} and \ref{branchb} were multiplied by other terms in the derivation above.

It has been assumed in the derivation that the operators commute with each other \cite{Ortigueira2015}. Alternative derivations of Eq.~\ref{semi} can be through stronger assumptions on $\L^n$ such as the existence of an inverse operator $\L^{-n}$, the Laplace transform (or its algebraic equivalents using operational calculus or by assuming harmonic solutions), and  electrical impedance methods.

A simple example is: $\L_a = a \D^n$, $\L_b = b \D^m$,  so that $\L_{eq} = \sqrt{a b}\ \D^{(n+m)/2}$ of order $(n+m)/2$. As mentioned before $\D^{(n+m)/2}$ can be explicitly defined. However, if $\L_a = a_1 + a_2 \D^n$ and $\L_b = b_1 + b_2 \D^m$, then \[
\L_{eq}^2 - \left[ a_1b_1 + a_2b_1 \D^n + a_1b_2 \D^m + a_2b_2 \D^{n+m} \right] = 0.
\]
No general explicit representation exists for $\L_{eq}$ which can, however, be written symbolically as \[ \L_{eq} = \left[ a_1b_1 + a_2b_1 \D^n + a_1b_2 \D^m + a_2b_2 \D^{n+m} \right]^{1/2} , \] which is also of equivalent order $(n+m)/2$.

\subsection{Infinitely multi-furcating trees}

\begin{figure}
\centering
\psfrag{p1}{$\phi_{in}$}
\psfrag{p2}{$\phi_{out}$}
\psfrag{La}{$\L_a$}
\psfrag{Lb}{$\L_b$}
\psfrag{L}{$\L_{eq}$}
\psfrag{m}{$m$ times}
\psfrag{n}{$n$ times}
\includegraphics[height=2in]{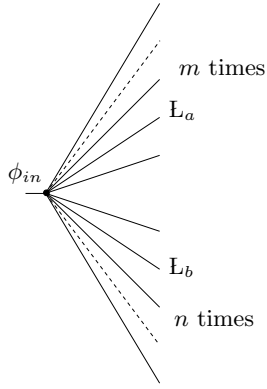}
\caption{First generation of multi-furcating tree network. Each
   junction has $m$ components with $\L_a$ and $n$ components with
   $\L_b$ connecting to a junction of a subsequent generation.} \label{ptree} \end{figure}

In an infinitely multi-furcating tree (called $p$-furcating in \cite{Mayes2011}), like the one in Fig.~\ref{ptree}, each junction has one connection from the lower generation and $n+m$ going to the subsequent generation, with $m$ of them with operators $\L_a$ and $n$ with $\L_b$. Following the same analysis as for the bifurcating tree except for $m$ times for $\L_a$ and $n$ times for $\L_b$, $\L_{eq}$ must satisfy \begin{multline*} \left( n+m-1 \right) \L_{eq}^2 + \left[ \left( m - 1 \right)
   \L_a + \left( n - 1 \right) \L_ b \right] \L_{eq} - \L_a \L_b \\ = \0.
\end{multline*}
This is of the form of Eq.~\ref{eq:quad1} with \begin{align*}
\B(\D) &= \frac{\left( m - 1 \right) \L_a + \left( n - 1 \right) \L_
   b}{ n+m-1} \\
\C(\D) &= - \frac{\L_a \L_b}{ n+m-1}.
\end{align*}
At least symbolically, the solution can be written as \begin{multline*}
\L_{eq} = \frac{1}{2 \left( n +   m - 1 \right)} \Big[ - \left( m  -1
     \right)   \L_a - \left( n  -1 \right) \L_ b  \\ + \left\{
     \left[ \left( m - 1 \right)   \L_a + \left( n - 1 \right) \L_ b 
\right]^2 + 4
     \left( n + m - 1 \right) \L_a \L_b \right\}^{1/2} \Big] \end{multline*} For a bifurcating tree, $n = m = 1$, which gives Eq.~\ref{semi} as before.

\subsection{Infinite ladders}

\begin{figure}
\psfrag{p1}{$\phi_{in}$}
\psfrag{p2}{$\phi_{out}$}
\psfrag{La}{$\L_a$}
\psfrag{Lb}{$\L_b$}
\includegraphics[width=3.25in]{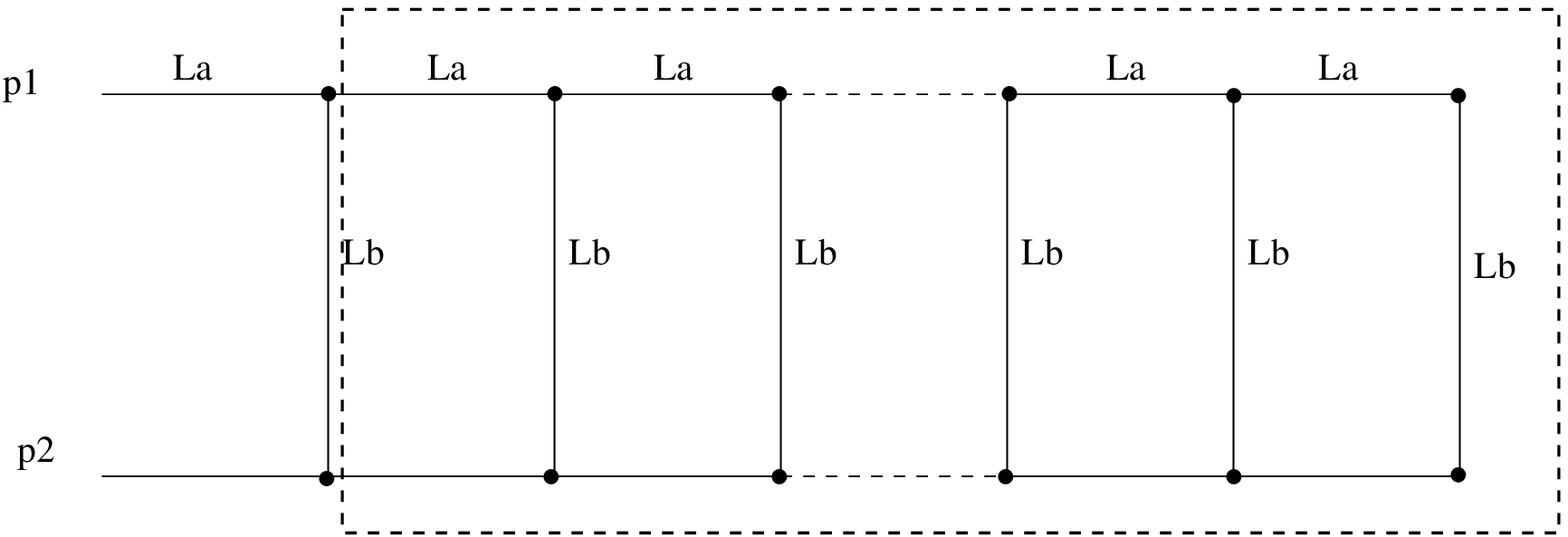}
\caption{Infinite ladder composed of operators $\L_a$ and
   $\L_b$; components in boxed area identical to whole.} \label{fig:ladder1} \end{figure}

Fig.~\ref{fig:ladder1} depicts an infinite ladder configuration \cite{Schiessel1993, Heymans1994}. Equations for the two paths and the equivalent network give \begin{subequations} \begin{align} \L_a (u_a) + \L_{eq} (u_a - u_b)  &= \phi_{in} - \phi_{out}, \label{lada}\\ \L_a (u_a) + \L_b (u_b)  &= \phi_{in} - \phi_{out}, \label{ladb}\\
\L_{eq}(u_a)  &=  \phi_{in} - \phi_{out} \label{equiv} \end{align} \end{subequations} The difference between Eqs.~\ref{lada} and \ref{ladb}, and between Eqs.~\ref{ladb} and Eq.~\ref{equiv} give \begin{subequations} \begin{align} (\L_{eq} + \L_b)(u_b)  &= \L_{eq} (u_a), \label{diffa}\\ (\L_{eq} - \L_a) (u_a)  &= \L_b(u_b) , \label{diffb} \end{align} \end{subequations} respectively. Operating on Eq.~\ref{diffb} by $(\L_{eq} + \L_b)$ and using Eq.~\ref{diffa} \begin{align*} (\L_{eq} + \L_b) \ (\L_{eq} - \L_a) (u_a) &= (\L_{eq} + \L_b) \
\L_b(u_b) , \\
&=\L_b \  \underbrace{(\L_{eq} + \L_b) (u_b)}_{= \L_{eq} (u_a)} .
\end{align*}
Canceling $u_a$ and reshuffling
\[
\L_{eq}^2 - \L_a \L_{eq} - \L_a \L_b = \0 , \] which is a quadratic in $\L_{eq}$ like Eq.~\ref{eq:quad1}.

\begin{figure}
\psfrag{p1}{$\phi_{in}$}
\psfrag{p2}{$\phi_{out}$}
\psfrag{La}{$\L_a$}
\psfrag{Lb}{$\L_b$}
\psfrag{L}{$\L_{eq}$}
\includegraphics[width=1in]{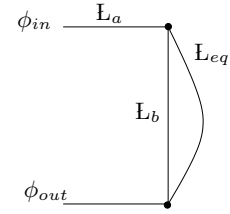}
\caption{Components in boxed area in Fig.~\ref{fig:ladder1} replaced
   by equivalent.}
\label{fig:ladder2}
\end{figure}

Taking, for example, $\L_a = a_1 + a_2 \D^n$ and $\L_b = b_1 + b_2 \D^m$ \[
\L_{eq}^2 - (a_1 + a_2 \D^n) \L_{eq} - (a_1 + a_2 \D^n) \ (b_1 + b_2
\D^m) = \0,
\]
which is of equivalent order $(m+n)/2$.

\section{Conclusions and Discussion}

Implicit operators $\L$ are defined as solutions of Eq.~\ref{def}. These operators are a generalization of commonly-used fractional-order derivatives $\D^\alpha$, where $\alpha$ is a real number. Implicit operators appear  in the modeling and analysis of complex systems that are composed of two infinitely self-repeating integer-order components, $\L_a$ and $\L_b$, of integer-order. It has been shown that the equivalent operator $\L_{eq}$ is implicit. The dynamic behavior of the system can be determined from $\L_{eq}$ alone. The derivation is very general and covers all potential-driven flows.

More generally, it is suggested that the response of complex systems can be studied by two equations, $\L(u) = h(t)$ and Eq.~\ref{def}, where the first governs the response of the system to an excitation $h(t)$, and the second defines the system operator $\L$. In the present work linear integro-differential operators appear at the component level, and a quadratic $\g(\L,\D)$ in $\L$ at a system level, but of course there are other possibilities depending on the application being considered. The independent variable has been taken to be time-like so, even though it has not been explicitly stated, the problems are initial value. Application to boundary value problems with a spatial independent variable is possible, as are extensions to partial differential equations. The future can be expected to bring both further analysis of the properties of implicit operators as well as diverse applications to physical problems.

Of course, for engineering applications, means to determine $\L_{eq}$ explicitly or to approximate it is important.  One approach would be to consider low- and high-frequency cases\cite{Kelly2009}, which can be used to make the operator explicit in the relevant frequency range. Other approaches could include Pad\'e approximations using rational functions, other types of series approximation, continued fractions, eigen-function expansions, etc. What approximation is best is an open question, and the answer could, of course, be application-specific.


\end{document}